\def\Lfour{L_4}
\def\Lfive{L_5}

\def\spose#1{\hbox to 0pt{#1\hss}}
\def\lta{\mathrel{\spose{\lower 3pt\hbox{$\sim$}}
    \raise 2.0pt\hbox{$<$}}}
\def\gta{\mathrel{\spose{\lower 3pt\hbox{$\sim$}}
    \raise 2.0pt\hbox{$>$}}}

\documentstyle[12pt,aasms4,epsf,rotate]{article}

\begin{document}

\title{Cartography for Martian Trojans}

\author{Serge Tabachnik and N. Wyn Evans}
\affil{Theoretical Physics, Department of Physics, 1 Keble Rd, Oxford,
OX1 3NP, UK}

\begin{abstract} 
The last few months have seen the discovery of a second Martian Trojan
(1998 VF31), as well as two further possible candidates (1998 QH56 and
1998 SD4).  Together with the previously discovered Martian satellite
5261 Eureka, these are the only known possible solar system Trojan
asteroids not associated with Jupiter. Here, maps of the locations of
the stable Trojan trajectories of Mars are presented. These are
constructed by integrating an ensemble of in-plane and inclined orbits
in the vicinity of the Martian Lagrange points for between 25 million
and 60 million years. The survivors occupy a band of inclinations
between $15^\circ$ and $40^\circ$ and longitudes between $240^\circ$
and $330^\circ$ at the $\Lfive$ Lagrange point.  Around the $\Lfour$
point, stable Trojans inhabit two bands of inclinations ($15^\circ < i
< 30^\circ$ and $32^\circ < i < 40^\circ$) with longitudes restricted
between $25^\circ$ and $120^\circ$.  Both 5261 Eureka and 1998 VF31
lie deep within one of the stable zones, which suggests they may be of
primordial origin.  Around Mars, the number of such undiscovered
primordial objects with sizes greater than 1 km may be as high as
$\sim 50$.  The two candidates 1998 QH56 and 1998 SD4 are not
presently on Trojan orbits and will enter the sphere of influence of
Mars within half a million years.
\end{abstract}


\section{INTRODUCTION}

The Lagrange points $\Lfour$ and $\Lfive$ are stable in the restricted
three body problem (e.g., Danby 1988). However, the long-term survival
of Trojans around the Lagrange points of the planets in the presence
of perturbations from the remainder of the Solar System is a difficult
and still unsolved problem (e.g., \'Erdi 1997). Jovian Trojan
asteroids have been known since the early years of this century, while
a number of Saturnian moons (e.g., Dione and Helene, Tethys and
Calypso, Tethys and Telesto) also form Trojan configurations with
their parent planet.  However, whether there exist Trojan-like bodies
associated with the other planets has been the matter of both
observational activity (e.g., Tombaugh 1961, Kowal 1971) and intense
theoretical speculation (e.g., Weissman \& Wetherill 1974, Mikkola \&
Innanen 1990, 1992).  The answer to this problem came in 1990, with
the discovery of 5261 Eureka, the first Trojan around Mars (see
Mikkola et al. 1994 for details). The last few months of 1998 have
seen further remarkable progress with the discovery of one certain
Martian Trojan, namely 1998 VF31, as well as two further candidates,
namely 1998 SD4 and 1998 QH56 (see the {\it Minor Planet Electronic
Circulars 1998-W04, 1998-R02, 1998-S20 } and the {\it Minor Planet
Circular 33085}). The suggestion that 1998 QH56 and 1998 VF31 might be
Martian Trojans was first made by G.V. Williams.

These recent discoveries raise very directly the following questions.
Are there any more Martian Trojans? If so, where should the
observational effort be concentrated? Of course, the first question
can only be answered at the telescope, but the resolution of the
second question is provided in this {\it Letter}. By integrating
numerically an ensemble of inclined and in-plane orbits in the
vicinity of the Martian Lagrange points for 25 and 60 million years
respectively, the stable r\'egimes are mapped out. On re-simulating
and sampling the ensemble of stable orbits, the probability density of
Martian Trojans as a function of longitude and inclination can be
readily obtained.  If a comparatively puny body such as Mars possesses
Trojans, the existence of such objects around the larger terrestrial
planets also merits very serious attention.  There are Trojan orbits
associated with Venus and the Earth that survive for tens of millions
of years (e.g., Tabachnik \& Evans 1998). If objects populating such
orbits exist, they must be small else they would have been found by
now.

\section{MARTIAN TROJANS}

Saha \& Tremaine (1992, 1994) have taken the symplectic integrators
developed by Wisdom \& Holman (1991) and added individual planetary
timesteps to provide a fast code that it is tailor-made for long
numerical integrations of low eccentricity orbits in a nearly
Keplerian force field. In our simulations, the model of the Solar
System consists of the eight planets from Mercury to Neptune, together
with test particles starting near the Lagrange points. The effect of
Pluto on the evolution of Martian Trojans is quite negligible. Of
course, the Trojan test particles are perturbed by the Sun and planets
but do not themselves exert any gravitational forces.  The initial
positions and velocities of the planets, as well as their masses, are
provided by the JPL Planetary and Lunar Ephemerides DE405 and the
starting epoch is JD 2440400.5 (28 June 1969).  All our simulations
include the most important post-Newtonian corrections, as well as the
effects of the Moon.  Individual timesteps are invaluable for this
work, as orbital periods are much smaller in the Inner Solar System
than the Outer.  For all the computations described in this Letter,
the timestep for Mercury is $14.27$ days. The timesteps of the planets
are in the ratio $1:2:2:4:8:8:64:64$ for Mercury moving outwards, so
that Neptune has a timestep of $2.5$ years. The Trojan particles all
have the same timestep as Mercury. These values were chosen after some
experimentation to ensure the relative energy error has a peak
amplitude of $\approx 10^{-6}$ over the tens of million year
integration timespans. After each timestep, the Trojan test particles
are examined to see whether their orbits have become hyperbolic or if
they have entered the planet's sphere of influence (defined as $r_{\rm
s} = a_{\rm p} M_{\rm p}^{2/5}$ where $a_{\rm p}$ and $M_{\rm p}$ are
the semimajor axis and mass of the planet). If so, they are
terminated. In methodology, our calculations are very similar to the
magisterial work on the Trojan problem for the four giant planets by
Holman \& Wisdom (1993). The earlier calculations of Mikkola \&
Innanen (1994, 1995) on the Trojans of Mars for timespans of between
tens of thousands and 6 million years have also proved influential.
Our integrations of Trojan orbits are pursued for durations ranging
from 25 to 60 million years, the longest integration periods currently
available. Nonetheless, the orbits have been followed for only a tiny
fraction of the age of the Solar System ($\sim 4.5$ Gigayears), so it
is wise to remain a little cautious about our results.

Figure 1 shows the results of our first experiment. Here, the orbits
of 1080 Trojan test particles around Mars are integrated for 25
million years.  The initial inclinations of the test particles (with
respect to the plane of Mars' orbit) are spaced every $2^\circ$ from
$0^\circ$ to $90^\circ$ and the initial longitudes (again with respect
to Mars) are spaced every $15^\circ$ from $0^\circ$ to
$360^\circ$. The starting semimajor axes and the eccentricities of the
Trojans are the same as the parent planet.  Only the test particles
surviving till the end of the 25 million year integration are marked
on the Figure.  The survivors occupy a band of inclinations between
$10^\circ$ and $40^\circ$ and longitudes between $30^\circ$ and
$120^\circ$ (the $\Lfour$ Lagrange point) or $240^\circ$ and
$330^\circ$ (the $\Lfive$ point).  On the basis of 4 million year
timespan integrations, Mikkola \& Innanen (1994) claim that stable
Martian Trojans have inclinations between $15^\circ$ and $30^\circ$
and between $32^\circ$ and $44^\circ$ with respect to Jupiter's orbit.
Our longer integrations seem to suggest a more complex
picture. Mikkola \& Innanen's instability strip between $30^\circ$ and
$32^\circ$ can be detected in Figure~1, but only for objects near
$\Lfour$ with initial longitudes $\lta 60^\circ$.  In particular, this
instability strip does not exist around $\Lfive$ and here Trojans with
starting inclinations $30^\circ < i < 32^\circ $ seem to be stable --
as is also evidenced by the recent discovery of 1998 VF31.  Marked on
the figure are the instantaneous positions of the two certain Martian
Trojans, namely 5261 Eureka (marked as a red circle) and 1998 VF31 (a
green circle), as well as the two candidates 1998 QH56 (a blue circle) and 1998
SD4 (a yellow circle). It is delightful to see that the two securely
established Trojans lie within the stable zone, which was computed by
Tabachnik \& Evans (1998) before the discovery of 1998 VF31.  In fact,
they live deep within the heart of the zone, suggesting that they may
even be primordial. The two candidates (1998 QH56 and 1998 SD4) lie
closer to the rim.  Let us finally note that Trojans starting off in
or near the plane of Mars' orbit are unstable. This has been confirmed
by an extensive survey of in-plane Martian Trojans. On integrating 792
test particles with vanishing inclination but with a range of
longitudes and semimajor axes, we found that all are unstable on
timescales of $60$ million years. Martian Trojans with low
inclinations are not expected.

It is useful to an observer hoping to discover further Trojans to
provide plots of the probability density. Accordingly, let us
re-simulate the stable zones with much greater resolution. This is
accomplished by placing a total of 746 test particles every $1^\circ$
in initial inclination and every $5^\circ$ in initial longitude so as
to span completely the stable regions. This ensemble of orbits is then
integrated and the orbital elements are sampled every 2.5 years to
provide the plots displayed in Figure 2. The upper panel shows the
meshed surface of the probability density as a function of both
inclination to the invariable plane and longitude with respect to the
planet. The asymmetry between the two Lagrange points is evident.  The
lower panels show the projections of the meshed surface onto the
principal planes -- in particular, for the inclination plot, we have
shown the contribution at each Lagrange point separately.  There are a
number of interesting conclusions to be drawn from the plots.  First,
as shown by the dotted line, the probability density is bimodal at
$\Lfour$. It possesses a flattish maximum at inclinations between
$15^\circ$ and $30^\circ$ and then falls sharply, before rising to a
second maximum at $36^\circ$. At $\Lfive$, all inclinations between
$15^\circ$ and $40^\circ$ carry a significant probability, though the
smaller inclinations in this band are most favored.  It is within
these inclination windows that the observational effort should be most
concentrated. Second, the probability density is peaked at longitudes
of $\sim 60^\circ$ ($\Lfour$) and $\sim 300^\circ$ ($\Lfive$). The
most likely place to observe one of these Trojans is indeed at the
classical locations of the Lagrange points. This is not intuitively
obvious, as any individual Trojan is most likely to be seen at the
turning points of its longitudinal libration. There are two reasons
why this effect is not evident in our probability density
plots. First, our figures refer to an ensemble of Trojans uniformly
populating the stable zone. So, the shape of the stable zone also
plays an important role in controlling the position of the maximum of
the probability density. Second, the positions of the Lagrange points
themselves are oscillating and so the turning points of the
longitudinal libration do not occur at the same locations, thus
smearing out the enhancement effect.

Table 1 lists the orbital elements of the two secure Martian Trojans
and the two candidates, as recorded by the Minor Planet Center. From
the instantaneous elements, it is straightforward to simulate the
trajectories of the objects.  Figure 3 shows the orbits plotted in the
plane of longitude (with respect to Mars) versus semimajor axis. As
the figures illustrate, both 5261 Eureka and 1998 VF31 are stable and
maintain their tadpole character (see e.g., Garfinkel 1977) for
durations of 50 million years. Based on preliminary orbital elements,
Mikkola et al. (1994) integrated the orbit of 5261 Eureka and found
that its longitudinal libration was large, quoting $297^\circ \pm
26^\circ$ as the typical range in the longitudinal angle.  Our orbit
of 5261 Eureka, based on the latest orbital elements, seems to show a
smaller libration of $285-314^\circ$. The remaining two objects that
have been suggested as Martian Trojans, 1998 QH56 and 1998 SD4, both
enter the sphere of influence of Mars -- in the former case after
$\sim 500\,000$ years, in the latter case after $\sim 100\,000$
years. Although the orbits are Mars crossing, their eccentricities
remain low and their inclinations oscillate tightly about mean values
until the Mars' sphere of influence is entered. It is possible that
these objects were once Trojans and have been ejected from the stable
zones, a possibility that receives some support from their locations
in Figure 1 at the fringes of the stable zones. Of course, another
possibility is that they are ejected asteroids from the Main Belt.

The fact that both confirmed Martian Trojans lie deep within the
stable zones in Figure 1 suggests that these objects may be
primordial. If so, we can get a crude estimate of possible numbers by
extrapolation from the number of Main Belt asteroids (c.f., Holman
1997, Evans \& Tabachnik 1999). The number of Main Belt asteroids
$N_{\rm MB}$ is $N_{\rm MB} \lta \Sigma_{\rm MB} A_{\rm MB} f$, where
$A_{\rm MB}$ is the area of the Main Belt, $\Sigma_{\rm MB}$ is the
surface density of the proto-planetary disk and $f$ is the fraction of
primordial objects that survive ejection (which we assume to be a
universal constant). Let us take the Main Belt to be centered on
$2.75$ AU with a width of $1.5$ AU.  The belt of Martian Trojans is
centered on $1.52$ AU and has a width of $ \lta 0.0025$ AU. If the
primordial surface density falls off inversely proportional to
distance, then the number of Martian Trojans $N_{\rm MT}$ is
\begin{equation}
N_{\rm MT} \lta \Bigl( {2.75\over 1.52} \Bigr) \Bigl( {1.52 \times 0.0025
\over 2.75 \times 1.5} \Bigr)  N_{\rm MB}
\approx 0.0016 N_{\rm MB}
\end{equation}
The number of known Main Belt asteroids with diameters $\gta 1$ km is
$\gta 40000$, which suggests that the number of Martian Trojans is
$\gta 50$.

\section{CONCLUSIONS}

Motivated by the recent discovery of a new Mars Trojans (1998 VF31) as
well as further possible candidates (1998 QH56, 1998 SD4), this paper
has provided maps of the stable zones for Martian Trojans and
estimates of the numbers of undiscovered objects. For Mars, the
observational effort should be concentrated at inclinations satisfying
$15^\circ < i < 30^\circ$ and $32^\circ < i < 40^\circ$ for the
$\Lfour$ Lagrange point and between $15^\circ$ and $40^\circ$ for
$\Lfive$.  These are the spots where the probability density is
significant (see Figure 2), though the lower inclinations in these
bands are slightly more favored than the higher.  Trojans in or close
the orbital plane of Mars are unstable.  Crude estimates suggest there
may be as many as $\sim 50$ undiscovered Martian Trojans with sizes
$\gta 1$ km . The orbits of 5261 Eureka and 1998 VF31 remain
Trojan-like for durations of at least 50 million years. The other
candidates, 1998 QH56 and 1998 SD4, are not currently Trojans, though
it is conceivable that they may once have been. Both objects will
probably enter the sphere of influence of Mars after $\lta 0.5$
million years.

\acknowledgments
NWE is supported by the Royal Society, while ST acknowledges financial
help from the European Community.  We wish to thank John Chambers,
Luke Dones, Seppo Mikkola, Prasenjit Saha and Scott Tremaine for many
helpful comments and suggestions. We are also grateful for the
remarkable service to the academic community provided by the Minor
Planet Center. The anonymous referee helpfully provided improved
orbital elements for the Trojan candidates for our integrations.

\eject

\begin{figure}
\caption{This figure shows the stability zones of the inclined
Trojans of Mars. The horizontal axis marks the longitude measured from
Mars and the vertical axis the inclination with respect to Mars of the
starting positions of test particles.  At outset, the array of
particles has inclinations spaced every $2^\circ$ and longitudes
spaced every $15^\circ$. The initial semimajor axes and eccentricities
of the Trojans are the same as Mars. Only the particles surviving till
the end of the 25 million year integration are marked on the figure,
which provides a map of the stable regions. All the objects starting
in-plane do not persist and only the inclined Trojans are stable. Also
marked on the figure are the instantaneous positions of the two
Martian Trojans, namely 5261 Eureka (marked as a red circle) and 1998 VF31
(a green circle), as well as the asteroids 1998 QH56 (a blue circle) and
1998 SD4 (a yellow circle).}
\end{figure}

\begin{figure}
\vspace*{0.5cm}
\caption{These figures show the most likely places to observe new 
Martian Trojans. They display the two-dimensional probability density
as a function of the inclination with respect to the invariable plane
and the longitude with respect to Mars (upper panel) together with the
projections onto the principal planes (lower panels). The figures are
constructed by re-simulating the stable regions displayed in Figure 1
at much greater resolution.  746 test particles are placed every
$1^\circ$ in inclination and every $5^\circ$ in longitude so as to
span the stable region and the trajectories are sampled every 2.5
years for $50\,000$ years. The overall normalisation of the probability
density is arbitrary. In the inclination plots, the contributions from
the Lagrange points are separated -- broken lines refer to $\Lfour$
and unbroken lines refer to $\Lfive$.}
\end{figure}

\begin{table}
\begin{center}
\begin{tabular}{|c|c|c|c|c|c|c|} \hline
Asteroid & $a$ & $e$ & $i$ [deg] & $\lambda$ [deg] & $H$ &
Diameter \\ \hline
5261 Eureka & 1.5235 & 0.0647 & 20.2802 & 301.4 & 16.1 & 2-4 km\\ \hline
1998 VF31 & 1.5241 & 0.1005 & 31.2966  & 290.9 & 17.1 & 1-2 km \\ \hline\hline
1998 QH56 & 1.5506 & 0.0307 & 32.2222 & 258.5 & 17.9 & 1-1.5 km \\ \hline
1998 SD4 & 1.5149 & 0.1254 & 13.6725 & 245.5 & 18.7 & 0.5-1.2 km \\ 
\hline
\end{tabular}
\end{center}
\vspace*{0.5cm}
\caption{This table lists some of the properties of the two definite
Martian Trojans, as well as two suggested candidates. These include
the instantaneous semimajor axis $a$, eccentricity $e$, inclination
from the J2000 plane $i$ and longitude measured from Mars $\lambda$. The
epoch is JD 2451200.5 (22 January 1999). The magnitude $H$ and the
approximate diameter of the object (inferred using albedos of
$0.05-0.25$) are also given. Most of this information is abstracted
from {\it Minor Planet Circulars 30250 and 33085} (Eureka and 1998
QH56) and {\it Minor Planet Electronic Circular 1998-W04 and 1998-S20}
(1998 VF31 and 1998 SD4).}
\end{table}

\begin{table}
\begin{center}
\begin{tabular}{|c|c|c|c|c|c|c|c} \hline
Asteroid & $\Delta a$ & $ \Delta e$ & $\Delta i$ [deg]& 
$\Delta \lambda$ [deg] &Lag. point &$D$ [deg] \\ \hline
5261 Eureka & $5.955\times 10^{-4}$& $0.08117$& $4.8874$& 
$29.47$ &${\Lfive}_{-15.2}^{14.3}$ &$12.0862$ \\ \hline
1998 VF31 &$1.386\times 10^{-3}$&$0.1128$ &$4.9722$ &$69.38$
&${\Lfive}_{-39.2}^{+30.1}$ &$47.523$ \\ 
\hline
\end{tabular}
\end{center}
\vspace*{0.5cm}
\caption{This table lists some of the properties of the orbits of the
two confirmed Martian Trojans, inferred from numerical
integrations. The table gives the maximum variation during the
entirety of the 50 million year integration timespan in the semimajor
axis $\Delta a$, in the eccentricity $\Delta e$, in the inclination
$\Delta i$ and in the longitude measured from Mars $\Delta
\lambda$. Both Trojans oscillate around the Lagrange point $\Lfive$ and
the superscript and subscript indicate the extent of the angular
libration. Part of this includes the oscillation of the Lagrange point
itself, so the final column $D$ is the peak to peak angular libration
measured from the Lagrange point.}
\end{table}

\begin{figure}
\caption{Plots of the longitude versus semimajor axis are shown for
the orbits of 5261 Eureka (upper panel) 1998 VF31 (lower panel). The
orbits are integrated for $50$ million years and are sampled every 
$10\,000$ years.}
\end{figure}

\eject

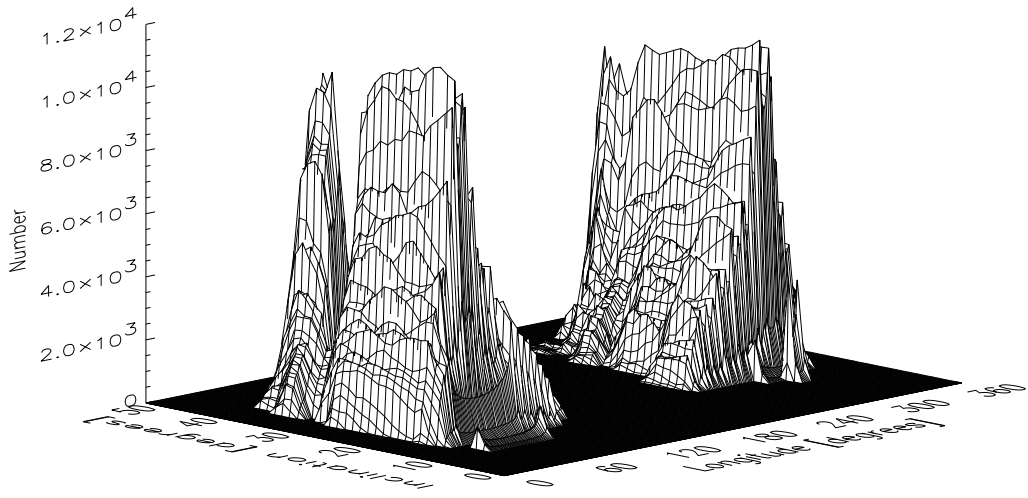
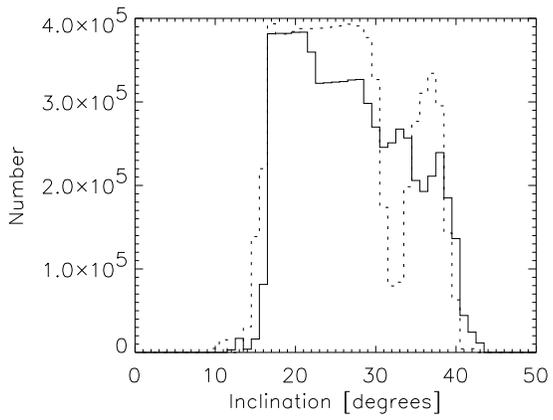
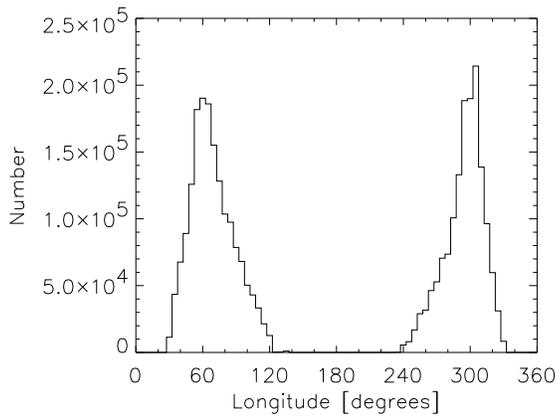
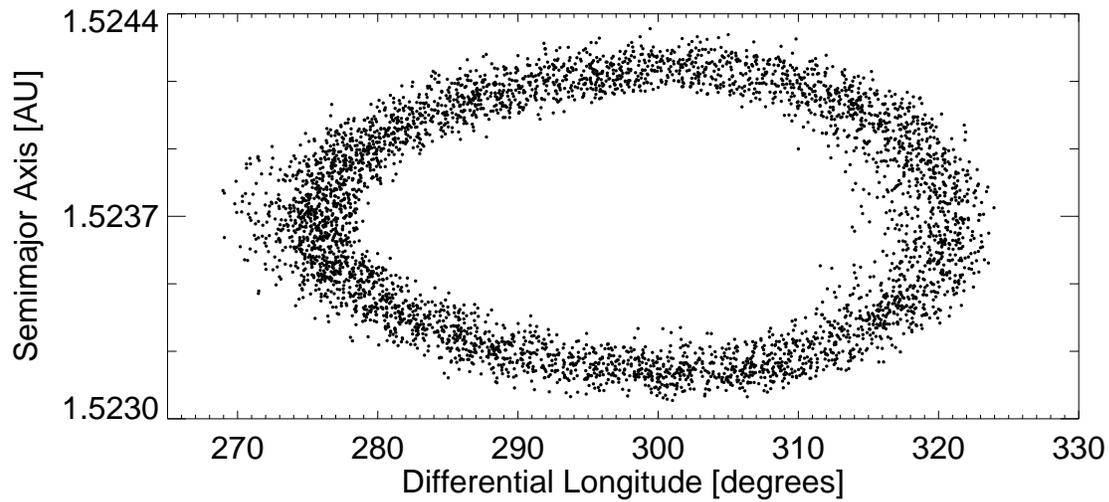
\end{document}